# Geometric Polarization–Virial Theorems for Field-Induced Dipolar Interactions and Force Reversal

Shuze Zhu

Center for X-Mechanics, Department of Engineering Mechanics, Zhejiang University, Hangzhou 310000, China

*To whom correspondence should be addressed. E-mail: shuzezhu@zju.edu.cn

**Abstract**

Electric fields can turn attraction into repulsion between far-separated atoms, yet the complete angular mechanical structure governing this reversal has remained unknown. Here, we uncover an entropy-like geometric structure: the cubically weighted interaction free energy, $r^3\mathcal{F}(r)$, is nondecreasing under physical dilation. This structure unifies two exact polarization–virial theorems for arbitrary stable quasistatic dipole assemblies and fully retarded dipole pairs under stated conditions. Consequently, the free energy has at most one zero and energetic cancellation precedes mechanical cancellation. A force zero retains strictly positive energy, whereas an energy zero retains an attractive force. The resulting analytical angular fine structure is resolved by a quantum harmonic oscillator pair, yielding the complete arbitrary-angle interaction energy and radial force that recover quantum-perturbative results. The conventional dipolar magic angle splits into strictly ordered energy- and force-zero angles; the latter uniquely bounds the radially repulsive sector and carries a positive energy barrier. Derived scaling laws and finite-sphere calculations provide experimentally testable signatures from microscopic dipoles to macroscopic polarizable bodies.

*Introduction.*— External electric fields redistribute polarization, tuning noncovalent interactions across scales[1–13]. Examples include enhanced water evaporation[11], our earlier demonstrations of activation-barrier reduction for both graphene origami[12] and polymer dispersion[13]. More recently, a series of arguments in Phys. Rev. Lett. [14–17] highlighted the intriguing physics of field-induced force reversal between two hydrogen atoms. These developments raise a fundamental question: can field-controlled attraction and repulsion be governed by a general mechanical principle rather than system-specific response?

The original Letter[14] reported a striking phenomenon in which a repulsive interaction emerges in the far zone between two hydrogen atoms in a static electric field. The subsequent Comment[15] emphasized that the leading sign-changing contribution is the electrostatic interaction between field-induced dipoles, while the Reply[16] retained an additional mixed one-photon interpretation. Further Comment[17] and subsequent analyses[18–20] clarified that the leading field-controlled far-zone force is associated with the field-induced electrostatic mean response, while the equilibrium fluctuation contribution retains the attractive Casimir-Polder character. These studies also identified additional field-dependent quantum corrections beyond the leading interaction. Across these communications, the conventional dipolar "magic angle", $\theta_c = \arccos(1/\sqrt{3})$, the zero of the leading fixed-dipole angular channel $1 - 3\cos^2\theta$ , marked the sign changes of both the leading-order force and energy.

But a zero of the leading channel need not be a zero of the complete interaction. Once mutual polarization and equilibrium fluctuations are included, the angular dependence of the energy and that of the force need not remain identical, so that $\theta_c$ may represent only one element of a finer mechanical structure. Although the above studies clarified the leading interaction, what remained lacking was a general arbitrary-angle formulation of the complete self-consistent interaction together with a firm theoretical foundation relating its mechanical derivative to the interaction free energy. What, then, is the complete mechanical structure of the self-consistent interaction? Where do the complete energy and force zeros occur, and how are they ordered with $\theta_c$? Does the force zero define a unique boundary between attraction and repulsion? More generally, is the observed force reversal an isolated feature of particular atomic models, or the manifestation of a broader and quantifiable mechanical structure? Such a structure would directly predict field-induced critical phenomena that are experimentally measurable.

To address these questions, here we establish two polarization-virial theorems that relate the complete interaction free energy to its mechanical virial [Fig. 1(a, b)]: one for general stable quasistatic dipole assemblies and the other for fully retarded reciprocal passive dipole pairs. Although the two theorems concern different response classes and mechanisms, a unified geometric structure arises: the cubically weighted interaction free energy, $r^3\mathcal{F}(r)$, is monotonic under physical dilation. In this sense, the theorem introduces an entropy-like arrow into interaction geometry: stable dilation selects a preferred direction for the evolution of the weighted free energy. Consequently, energetic cancellation precedes mechanical cancellation: the force zero of the complete field-controlled interaction retains positive free energy, while an energy zero retains an attractive force. We then realize the resulting angular fine structure by an exactly solvable quantum harmonic oscillator pair. Quantum perturbation theory identifies the angle-resolved electrostatic, mutual-polarization, and equilibrium-fluctuation sectors, while a direct Gaussian solution of the same Hamiltonian yields the complete arbitrary-angle energy and radial force. The conventional magic angle $\theta_c$ thereby splits into distinct, strictly ordered energy- and force-zero angles, with the force zero forming the unique boundary of the repulsive angular sector and carrying a positive interaction-energy barrier [Fig. 1(c)]. Finally, we derive scaling laws for critical fields, cancellation distances, angular shifts, and barrier heights, elevating field-induced force reversal from a model-specific observation to a unified and experimentally testable mechanical theory. More broadly, the persistence of ordered cancellation in finite-sphere calculations strongly suggests that the distinction between energetic and mechanical cancellation could survive across scales—from microscopic dipoles to macroscopic polarizable bodies.

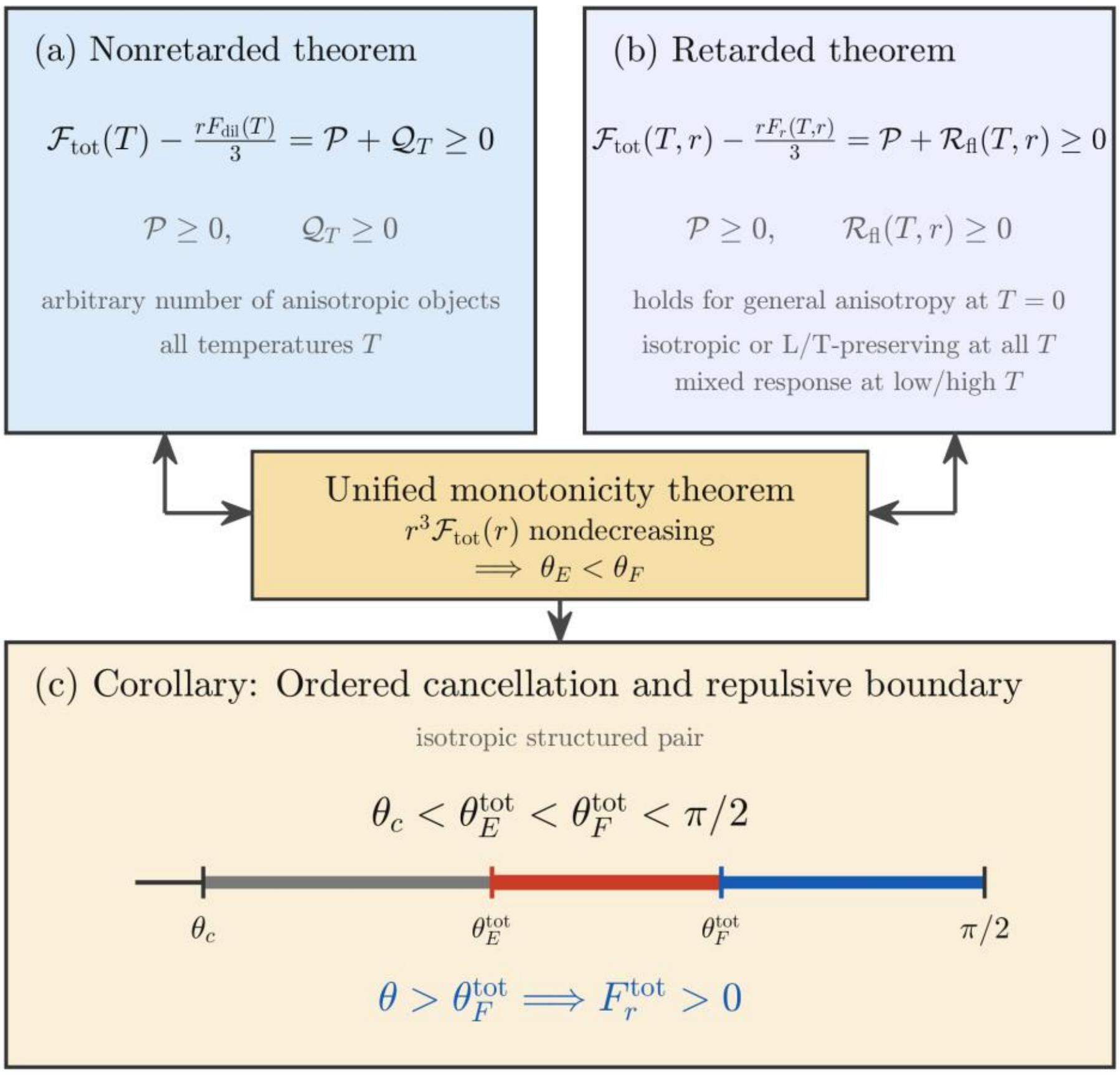

FIG. 1 Scope and consequences of the polarization-virial theorems. (a) The nonretarded theorem applies to arbitrary number of general anisotropic objects at all temperature. (b) The retarded matrix theorem applies to arbitrary anisotropic response at zero temperature. All-temperature extensions apply to isotropic or longitudinal-transverse-preserving response, while general mixed response is covered at sufficiently low and high temperature. The two theorems are unified by a common entropy-like geometric structure: the cubically weighted interaction free energy evolves in only one direction under physical dilation. (c) Observable consequences of ordered cancellations and repulsive boundary for isotropic structured pair. This observation persists in a finite-sphere calculation, extending the scope beyond the point-dipole model.

*General quasistatic polarization-virial theorem.*— Consider $N$ linearly polarizable electric or magnetic objects. Let $\boldsymbol{d} \in \mathbb{R}^{3N}$ collect their induced dipoles, $\boldsymbol{D} > 0$ denote the block-diagonal isolated static response, $\boldsymbol{s}$ the imposed source vector (e.g., an externally applied field sampled at the object locations), and let $\boldsymbol{A}(r) = \boldsymbol{A}^{\mathrm{T}}(r)$ the intersite interaction matrix. The fixed-source quadratic functional is $U(\boldsymbol{d}; r) = \frac{1}{2}\boldsymbol{d}^{\mathrm{T}}\boldsymbol{D}^{-1}\boldsymbol{d} - \boldsymbol{s}^{\mathrm{T}}\boldsymbol{d} + \frac{1}{2}\boldsymbol{d}^{\mathrm{T}}\boldsymbol{A}(r)\boldsymbol{d}$, comprising the polarization self-energy (relative to the isolated reference), source–dipole coupling, and pairwise dipole–dipole interaction terms. The minimization of $U(\boldsymbol{d}; r)$ gives $\boldsymbol{d} = [\boldsymbol{D}^{-1} + \boldsymbol{A}]^{-1}\boldsymbol{s}$. The isolated response (letting $\boldsymbol{A} = \boldsymbol{0}$) is $\boldsymbol{d}_0 = \boldsymbol{D}\boldsymbol{s}$, and the quasistatic interaction energy after subtraction of the isolated reference is $V_{\mathrm{stat}} = \frac{1}{2}\boldsymbol{s}^{\mathrm{T}}(\boldsymbol{D} - [\boldsymbol{D}^{-1} + \boldsymbol{A}]^{-1})\boldsymbol{s}$. At fixed geometry, we introduce an auxiliary amplitude parameter $\lambda$ to control the interaction strength $\boldsymbol{A} \mapsto \lambda\boldsymbol{A}$ while holding $\boldsymbol{D}$ and $\boldsymbol{s}$ fixed, so that the interaction energy $V_{stat}(\lambda)$ follow continuously from the isolated response at $\lambda = 0$ to the physical system at $\lambda = 1$. By the Hellmann–Feynman theorem, $\partial_\lambda V_{stat}(\lambda) = \frac{1}{2}\boldsymbol{d}_\lambda^{\mathrm{T}}\boldsymbol{A}\,\boldsymbol{d}_\lambda$, which remarkably simplifies $V_{stat} - \partial_\lambda V_{stat}$ to a non-negative quadratic form:

$$V_{\mathrm{stat}} - \partial_{\ln\lambda}V_{\mathrm{stat}}|_{\lambda=1} = \frac{1}{2}(\boldsymbol{d}_0 - \boldsymbol{d})^{\mathrm{T}}\boldsymbol{D}^{-1}(\boldsymbol{d}_0 - \boldsymbol{d}) \equiv \mathcal{P} \geq 0. \tag{1}$$

The remainder $\mathcal{P}$ measures the feedback-induced change of the induced moments from their isolated values.

For a fixed-shape dilation, $F_{\mathrm{dil}}^{\mathrm{stat}} \equiv -\partial_r V_{\mathrm{stat}}$ is the force conjugate to this collective dilation. Simple algebra gives at $F_{\mathrm{dil}}^{\mathrm{stat}} = \frac{3}{r}\partial_{\ln\lambda}V_{\mathrm{stat}}|_{\lambda=1}$, Therefore, for the above static mean responses:

$$V_{\mathrm{stat}} - \frac{r\,F_{\mathrm{dil}}^{\mathrm{stat}}}{3} = \mathcal{P} \geq 0. \tag{2}$$

This theorem applies to arbitrary particle number, unequal objects, anisotropic positive-definite response tensors, and arbitrary fixed dimensionless geometry on a stable branch. For a many-body configuration, $rF_{dil}^{stat} = \sum_i r_i \cdot F_i$ is the configurational virial; for a pair it reduces to the ordinary radial virial.

Within the linear-response model considered here, the total interaction free energy at finite temperature factorizes as $\mathcal{F}_{total}(T) = V_{\mathrm{stat}} + \mathcal{F}_{fl}(T)$ , where $\mathcal{F}_{fl}(T)$ is the equilibrium fluctuation contribution. Scaling the complete dynamic interaction by $\lambda$ produces a nonnegative fluctuation remainder. For a fully nonretarded Gaussian system, the same algebraic structure yields the nonretarded Gaussian theorem at all temperatures:

$$\mathcal{F}_{total}(T) - \frac{r\,F_{\mathrm{dil}}(T)}{3} = \mathcal{P} + \mathcal{Q}_T \geq 0, \tag{3}$$

where $\mathcal{Q}_T \geq 0$ is the nonnegative fluctuation remainder. Theorem Eq. (2) and (3) follow from the common-amplitude identity in Eq. (1). Retardation breaks its direct identification with the physical radial derivative through the $\xi r/c$ dependence of the propagation kernel, requiring a separate treatment.

*Retarded polarization-virial theorem.*— This theorem applies to two electric-dipole objects. We first present the zero-temperature result. The key insight is that, although the frequency-resolved fluctuation remainder can be negative, its integral over the full imaginary-frequency spectrum is nonnegative. The proof combines the round-trip expansion of the determinant with the matrix Stieltjes representation implied by passivity and causality, using the following standard ingredients: (i) free-space propagation kernel is explicitly known and has the scaling form $\mathbf{\Gamma}(i\xi, r) = r^{-3}\hat{\mathbf{\Gamma}}(\xi r/c)$; (ii) isolated imaginary-frequency response tensor $\boldsymbol{\alpha}_i(i\xi)$ for each dipole; (iii) normalized single-pass and round-trip matrices $\boldsymbol{X} = \boldsymbol{\alpha}_1^{1/2}\mathbf{\Gamma}\boldsymbol{\alpha}_2^{1/2}$ and $\boldsymbol{Y} = \boldsymbol{X}\boldsymbol{X}^{\mathrm{T}} \succcurlyeq 0$; (iv) zero-temperature fluctuation interaction energy, given by the nonperturbative coupled-dipole determinant[21] $V_{fl}(0, r) = \frac{\hbar}{2\pi}\int_0^\infty d\xi \ln \det[\boldsymbol{I} - \boldsymbol{Y}(\xi, r)]$. The static fixed-source mean sector yields a separate nonnegative remainder $\mathcal{P}$ provided a finite positive-definite zero-frequency response $\boldsymbol{\alpha}_i(0) > 0$.

The proof requires the uniform stability margin and absolute convergence assumptions which justify the round-trip expansion $\ln\det(\boldsymbol{I} - \boldsymbol{Y}) = -\sum_{n=1}^{\infty}\frac{1}{n}\mathrm{Tr}(\boldsymbol{Y}^{\boldsymbol{n}}) = -\sum_{n=1}^{\infty}\frac{t_n}{n}$ where $t_n(\xi, r) \equiv \mathrm{Tr}[\boldsymbol{Y}(\xi, r)^n] \geq 0$.Defining $s_n(\xi, r) \equiv -\partial_{\ln\xi} t_n(\xi, r)\big|_{response}$ as the contribution from differentiating the isolated response matrices, the Stieltjes representation of the passive response gives the bound $0 \leq s_n(\xi, r) \leq 4n\, t_n(\xi, r)$, so that the inequality $\left(6 - \frac{2}{n}\right)t_n - \frac{s_n}{n} \geq 2\left(1 - \frac{1}{n}\right)t_n \geq 0$ holds. Remarkably, after inserting the round-trip expansion and performing integration by parts over continuous imaginary-frequency spectrum, boundary terms vanish so that the fluctuation remainder $\mathcal{R}_{fl}$, defined as $\mathcal{R}_{fl}(0, r) \equiv V_{fl}(0, r) - \frac{rF_{fl}(0,r)}{3}$, becomes $\mathcal{R}_{fl}(0, r) = \frac{\hbar}{6\pi}\sum_{n=1}^{\infty}\int_0^\infty d\xi\left[\left(6 - \frac{2}{n}\right)t_n - \frac{s_n}{n}\right] \geq 0$ .

Combining $\mathcal{P}$ and $\mathcal{R}_{fl}$ gives the zero-temperature general matrix theorem:

$$V_{tot}(0, r) - \frac{rF_r(0, r)}{3} = \mathcal{P} + \mathcal{R}_{fl}(0, r) \geq 0. \tag{4}$$

The theorem applies to two reciprocal passive electric-dipole objects, not necessarily identical, with arbitrary fixed relative orientations and fully retarded, nonperturbative round-trip coupling. Under the same assumptions, the inequality is strict if $\mathcal{P} > 0$, or if $\boldsymbol{Y}(\xi, r) \neq \mathbf{0}$ on a set of frequencies of positive measure.

Its finite-temperature extension has three levels. First, for two passive isotropic electric-dipole objects, not necessarily identical, and more generally for anisotropic responses preserving the longitudinal and transverse subspaces defined by the separation direction, every stable Matsubara channel contributes a nonnegative radial remainder. Therefore, for every temperature:

$$\mathcal{F}_{tot}(T, r) - \frac{rF_r(T, r)}{3} = \mathcal{P} + \mathcal{R}_{fl}(T, r) \geq 0. \tag{5}$$

Second, for general anisotropic matrix response with longitudinal-transverse mixing, Eq. (5) holds at sufficiently low temperatures, provided the zero-temperature remainder is strict and the Matsubara sum converges regularly, and at sufficiently high temperatures, when a nonzero static round-trip coupling makes the positive zero-frequency contribution dominate the exponentially suppressed nonzero-frequency tail. Third, no unified positivity theorem is asserted at intermediate temperatures: the integration-by-parts identity for the continuous frequency integral has no direct counterpart for a discrete Matsubara sum.

*The unified monotonicity theorem and ordered-cancellation corollaries.*— The geometry of polarization-virial theorems can be visualized through the cubically weighted interaction $W_3[V](r) \equiv r^3 V(r)$. Since radial force $F_r(r) = -\partial_r V(r)$, we have $\mathrm{d}W_3[V]/\mathrm{d}r = 3r^2 V + r^3 \mathrm{d}V/\mathrm{d}r = 3r^2(V - rF_r/3)$. Thus:

$$V - \frac{rF_r}{3} \geq 0 \iff \frac{\mathrm{d}}{\mathrm{d}r}[r^3 V(r)] \geq 0. \tag{6}$$

For a quasistatic pair, the theorem gives $\mathrm{d}[r^3 V_{stat}]/\mathrm{d}r = 3r^2 \mathcal{P}(r) \geq 0$. For the complete retarded interaction, the corresponding relations are $\mathrm{d}[r^3 V_{tot}(0,r)]/\mathrm{d}r = 3r^2[\mathcal{P}(r) + \mathcal{R}_{fl}(0,r)] \geq 0$ and $\mathrm{d}[r^3 \mathcal{F}_{tot}(T,r)]/\mathrm{d}r = 3r^2[\mathcal{P}(r) + \mathcal{R}_{fl}(T,r)] \geq 0$. For a many-body configuration, $r$ denotes the collective dilation parameter that defines the configurational virial.

Hence the cubically weighted interaction is nondecreasing under physical dilation on every domain where the corresponding polarization-virial theorem applies. If the remainder is strictly positive throughout that domain, it is strictly increasing and the interaction has at most one energy zero. Whenever the strict theorem applies,

$$\mathcal{F}_{tot} = 0 \Longrightarrow F_r < 0, \qquad F_r = 0 \Longrightarrow \mathcal{F}_{tot} > 0 \tag{7}$$

An energy zero therefore retains an attractive radial force, whereas a force zero retains strictly positive interaction energy. Hence energetic and mechanical cancellation cannot coincide.

Consider a separation scan at fixed temperature, field strength, and field orientation. Assume that the complete interaction has an energy zero $r_E$ and any force zero $r_{F,j}$ so that $\mathcal{F}_{tot}(r_E) = 0$, $F_r(r_{F,j}) = 0$. At energy zero, $W_3[\mathcal{F}_{tot}](r_E) = 0$. At every force zero where the total remainder is strict, the polarization-virial theorem gives $W_3[\mathcal{F}_{tot}](r_{F,j}) = r_{F,j}^3 \mathcal{F}_{tot}(r_{F,j}) > 0$. Because $W_3[\mathcal{F}_{tot}]$ is nondecreasing, we obtain $r_E < r_{F,j}$. This ordering is the nontrivial consequence of the polarization-virial constraint: energetic cancellation necessarily occurs before mechanical cancellation, even if multiple force zeros exist.

Consider an angular scan at fixed temperature, field strength, and separation. Let $\theta$ be the angle between the uniform external static field and the separation vector. For a structured isotropic pair, the complete radial force is strictly increasing with $\theta$. If $\mathcal{F}_{tot}(\theta_E) = 0$, then the polarization-virial theorem gives $F_r(\theta_E) < 0$. Therefore, whenever an interior total-force zero $\theta_F$ exists, $\theta_E < \theta_F$, and $F_r(\theta) > 0$ for $\theta > \theta_F$ so that $\theta_F$ is the unique boundary of the radially repulsive angular sector.

*Angular geometry of force reversal and magic-angle splitting.*— The axial configurations (i.e., $\theta = \pi/2$ or $\theta = 0$) considered in the recent Phys. Rev. Lett. arguments[14–17] do not exhaust the mechanical structure of the force-reversal problem. Here we reveal its general angular structure through an exact realization of the polarization-virial theorems. We solve exactly, and nonperturbatively, the coupled system of two identical isotropic quantum harmonic oscillators in a uniform static field at arbitrary angle. Because the Hamiltonian is quadratic, its exact solution separates the interaction into a field-dependent mean contribution and a field-independent equilibrium fluctuation contribution. The perturbative expansion of the same solution resolves the field-induced electrostatic, mutual-polarization, and London sectors, whereas the normal-mode solution resums the entire series exactly. This factorized solution provides a direct realization of the polarization-virial constraint on the complete interaction, rather than on either sector separately. Consequently, the complete energy- and force-zero angles split and obey a strict ordering, with the conventional magic angle recovered only in the weak-coupling limit.

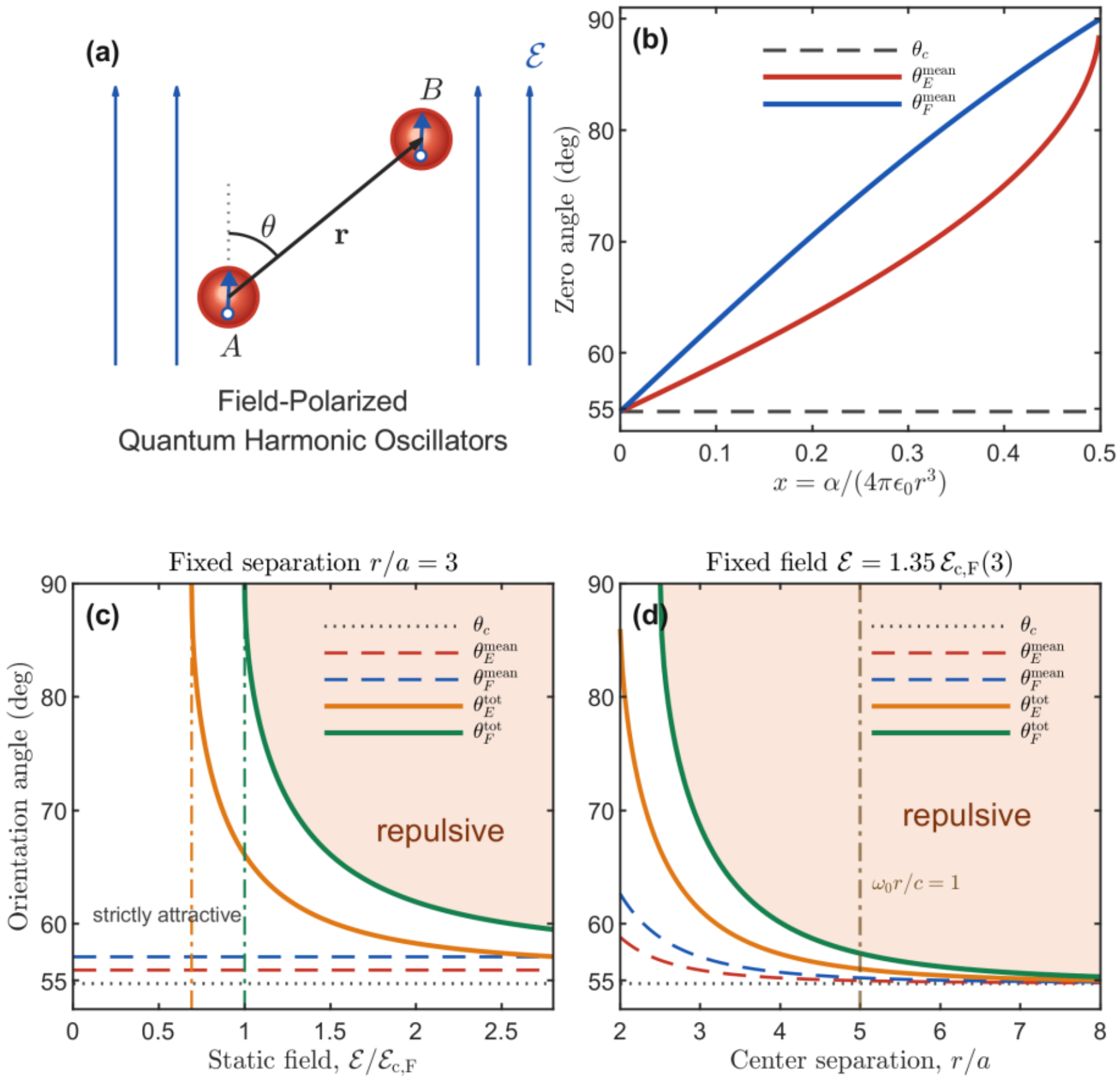

FIG. 2. General angular geometry of field-induced force reversal and magic-angle splitting. (a) Schematic of two field-polarized quantum harmonic oscillators, representing neutral polarizable atoms. The separation vector $\boldsymbol{r}$ forms an angle $\theta$ with the applied static field $\mathcal{E}$ .(b) Static energy- and force-zero angles, $\theta_E^{mean}$ and $\theta_F^{mean}$, as functions of the dimensionless coupling strength $x$; the dashed line denotes the conventional fixed-dipole magic angle $\theta_c$ . (c) Retarded fixed-separation scan at $r/a = 3$, with the field normalized by the critical field strength $\mathcal{E}_{c,F}$ for force reversal. (d) Retarded fixed-field separation scan at $\mathcal{E} = 1.35\mathcal{E}_{c,F}(r/a = 3)$; the vertical line marks the retardation crossover $\frac{\omega_0 r}{c} = 1$. Solid curves denote the total energy and force zeros, dashed curves the static mean-sector zeros, and dotted curves the conventional magic angle. The shaded regions indicate the net-radially-repulsive sector above the total-force zero.

Two identical neutral atoms are modelled as three-dimensional isotropic quantum harmonic oscillators separated by $r$. A uniform static electric field $\mathcal{E}$ forms an angle $\theta$ with separation axis [Fig. 2(a)]. In this realization, the fixed-source interaction $V_{\rm stat}$ of the quasistatic theorem is precisely the source-driven mean contribution $V_{\rm mean}$ of the displaced Gaussian state. The quadratic Hamiltonian is solved by introducing symmetric and antisymmetric normal coordinates for each Cartesian component. The interatomic coupling then shifts the two normal-mode frequencies, while the uniform field drives only the symmetric mode. Completing the square in that mode precisely separates the field-dependent mean energy from the field-independent fluctuation energy. Thus, $V_{\rm tot}(r,\theta) = V_{\rm mean}(r,\theta) + V_{\rm fl}(r)$, with all higher orders in the dipole coupling retained. Expanding this exact result reproduces the leading electrostatic, mutual-polarization, and London contributions independently identified by angle-resolved quantum perturbation theory (Appendix).

Define $x = \frac{\alpha(0)}{4\pi\varepsilon_0 r^3}, 0 < x < \frac{1}{2}$, where $\alpha(0)$ is the isolated static polarizability. The upper bound of $x$ is the stability condition. The exact mean interaction energy is $V_{mean}(r,\theta) = V_\perp(r)\sin^2\theta + V_\parallel(r)\cos^2\theta = \alpha(0)\,\mathcal{E}^2 x\left[\frac{\sin^2\theta}{1+x} - \frac{2\cos^2\theta}{1-2x}\right]$, whose corresponding radial force is $F_r^{\rm mean}(r,\theta) = F_\perp(r)\sin^2\theta + F_\parallel(r)\cos^2\theta = \frac{3\alpha(0)\mathcal{E}^2 x}{r}\left[\frac{\sin^2\theta}{(1+x)^2} - \frac{2\cos^2\theta}{(1-2x)^2}\right]$. They satisfy the exact static polarization-virial identity $V_{mean} - \frac{rF_r^{\rm mean}}{3} = \mathcal{P}$ where $\mathcal{P} = \alpha(0)\,\mathcal{E}^2 x^2\left[\frac{\sin^2\theta}{(1+x)^2} + \frac{4\cos^2\theta}{(1-2x)^2}\right] \geq 0$. Note that $V_\perp(r), F_\perp(r) > 0$ and $V_\parallel(r), F_\parallel(r) < 0$, so that the transverse field component generates a repulsive channel.

For $x \ll 1$, the exact mean interaction expands as $V_{mean}(r,\theta) = \frac{\alpha^2(0)\mathcal{E}^2}{4\pi\epsilon_0 r^3}(1 - 3\cos^2\theta) - \frac{\alpha^3(0)\mathcal{E}^2}{(4\pi\epsilon_0)^2 r^6}(1 + 3\cos^2\theta) + O(r^{-9})$. Thus, the angular structure of the total interaction is $V_{\rm tot}(r,\theta) = V_{el}(r,\theta) + V_{pol}(r,\theta) + V_{\rm fl}(r)$. Here $V_{\rm el}(r,\theta) = \frac{\alpha^2(0)\mathcal{E}^2}{4\pi\epsilon_0 r^3}(1 - 3\cos^2\theta)$ is temperature-independent interaction between the leading field-induced mean dipoles, identified in previous Comment[15]; $V_{\rm pol}(r,\theta) = -\frac{\alpha^3(0)\mathcal{E}^2}{(4\pi\epsilon_0)^2 r^6}(1 + 3\cos^2\theta)$ is the temperature-independent mutual-polarization correction; $V_{\rm fl}(r) = -\frac{3\hbar\omega_0\alpha^2(0)}{4(4\pi\epsilon_0)^2 r^6} + O(r^{-12})$ is the zero-temperature equilibrium fluctuation interaction. Only the sign of $V_{\rm el}(r,\theta)$ can change at $\theta_c = \arccos(1/\sqrt{3})$. Both $V_{\rm pol}(r,\theta)$ and angle-independent $V_{\rm fl}(r)$ always produce attractive forces.

For the retarded regime, integrating out the electromagnetic field gives the zero-temperature fluctuation interaction $V_{fl}^{ret}(0,r) = \frac{\hbar}{2\pi}\int_0^\infty d\xi[\ln(1-q_L) + 2\ln(1-q_T)]$ where $q_L = \alpha^2(i\xi)\Gamma_L^2(i\xi,r)$ and $q_T = \alpha^2(i\xi)\Gamma_T^2(i\xi,r)$. Here $\alpha(i\xi)$ is the scalar dynamic polarizability of either identical isotropic atom evaluated at imaginary frequency $i\xi$, while $\Gamma_L$ and $\Gamma_T$ are the longitudinal and transverse eigenvalues of free-space dipole-interaction kernel. Thus $V_{\rm tot} = V_{\rm mean} + V_{fl}^{ret}$. On the stable branch, $0 \leq q_L, q_T < 1$, so that $V_{fl}^{ret}(0,r) < 0$. By the polarization-virial theorem, the fluctuation interaction force must be attractive (negative) whenever the coupling is nonzero. Because $q_L$ and $q_T$ do not depend on $\theta$, $V_{fl}^{ret}$ cannot itself generate an angular force reversal.

In the retarded far zone, the leading asymptotic electrostatic force (temperature-independent) is $F_r^{el}(r,\theta) \sim \frac{3\alpha^2(0)\mathcal{E}^2}{4\pi\varepsilon_0 r^4}(1-3\cos^2\theta)$ and retard fluctuation force are $F_r^{fl,ret}(0,r) \sim -7C_7 r^{-8}$ where $C_7 > 0$. Therefore, for $\theta > \theta_c$, the leading field-controlled force is the repulsive electrostatic force between the field-induced mean dipoles. The retarded equilibrium fluctuation force remains attractive and decays more rapidly with distance.

The above microscopic realization also determines the geometry of complete force reversal. The dependence structures of total free energy and total force are $\mathcal{F}_{\text{tot}}(r,T,\theta,\mathcal{E}) = V_{mean}(r,\theta,\mathcal{E}) + \mathcal{F}_{\text{fl}}(r,T)$ and $F_{\text{r}}^{tot}(r,T,\theta,\mathcal{E}) = F_r^{mean}(r,\theta,\mathcal{E}) + F_{\text{r}}^{fl}(r,T)$ under applicable temperature classes. Thus $\frac{\partial F_r^{tot}}{\partial\theta} = \frac{\partial F_r^{mean}}{\partial\theta} = \frac{6\alpha(0)\mathcal{E}^2 x}{r}\sin\theta\cos\theta\left[\frac{1}{(1+x)^2}+\frac{2}{(1-2x)^2}\right] > 0$ for $0 < \theta < \pi/2$. Therefore, previous *Corollaries* guarantee the ordering of total energy and force zeros as $\theta_E^{\text{tot}} < \theta_F^{\text{tot}}$. From our exact solution, these critical angles are analytical as $\cos^2\theta_E^{tot} = \left(\frac{1-2x}{3}\right)\left[1-\frac{\mathcal{E}_{c,E}^2(r,T,\pi/2)}{\mathcal{E}^2}\right]$, $\cos^2\theta_F^{tot} = \frac{(1-2x)^2}{3+6x^2}\left[1-\frac{\mathcal{E}_{c,F}^2(r,T,\pi/2)}{\mathcal{E}^2}\right]$, where $\mathcal{E}_{c,E}^2(r,T,\pi/2) = -\frac{(1+x)\mathcal{F}_{fl}(r,T)}{\alpha(0)x}$ and $\mathcal{E}_{c,F}^2(r,T,\pi/2) = -\frac{r(1+x)^2F_r^{fl}(r,T)}{3\alpha(0)x}$ reveal the critical field strength associated with energy zero or force reversal at $\theta = \frac{\pi}{2}$. When $\mathcal{E}\to\infty$, $\cos^2\theta_E^{tot} \to \cos^2\theta_E^{mean} = \frac{1-2x}{3}$ and $\cos^2\theta_F^{tot} \to \cos^2\theta_F^{mean} = \frac{(1-2x)^2}{3+6x^2}$, which are the exact mean-sector energy and force zeros. Therefore, a sequence of zero ordering holds as $\theta_c < \theta_E^{mean} < \{\theta_F^{mean}, \theta_E^{tot}\} < \theta_F^{\text{tot}} < \pi/2$. Note that $\theta_E^{tot} \to \theta_E^{mean} < \theta_F^{mean}$ at $\mathcal{E}\to\infty$, whereas sufficiently close to the energy threshold one may have $\theta_E^{tot} \to \pi/2 > \theta_F^{mean}$. If $F_r^{tot}(r,T,\pi/2) > 0$, then $\theta_F^{\text{tot}}$ is the unique boundary of net-radially-repulsive angular sector:

$$\theta > \theta_F^{\text{tot}} \implies F_r^{tot}(r,T,\theta) > 0. \tag{8}$$

Fig. 2(b) shows the static mean-sector ordering, whereas Fig. 2(c-d) show the corresponding total-zero curves and the repulsive boundary after the fluctuation is included in a retarded scan, within $(\mathcal{E},\theta)$ or $(r,\theta)$ planes at zero temperature. The ordered cancellation Corollary (unique boundary) ensures that the general repulsive regime is a continuous region bound by a hypersurface $F_r^{tot}(\mathcal{E},r,T,\theta) = 0$.

*Scaling laws of critical phenomena and experimental observables.*— Questions like what is the critical force-reversal field strength at given $\theta$ intrigue experimentalists. The polarization-virial theorems establish the ordering of zeros and sign of energy and forces without assuming a particular asymptotic form for the fluctuation interaction. Explicit scaling laws require the controlled far-zone reduction $\mathcal{F}_{tot}(r) \sim \frac{C_3(\theta)\mathcal{E}^2}{r^3} - \frac{C_q(T)}{r^q}, q > 3$, where $C_3(\theta) = \frac{\alpha^2(0)}{4\pi\epsilon_0}(1-3\cos^2\theta) > 0$ and $C_q(T) > 0$. The first term is the leading electrostatic interaction between the field-induced mean dipoles, and the second is an attractive fluctuation background. Balancing these contributions gives a series of experimentally testable scaling laws. For example, at fixed $(r, T, \theta)$, the force-reversal critical field strength obeys $\mathcal{E}_{c,F} \propto C_q^{\frac{1}{2}} r^{\frac{3-q}{2}} (1 - 3\cos^2\theta)^{-\frac{1}{2}}$. At fixed $(\mathcal{E}, T, \theta)$, the corresponding force-zero separation scales as $r_{c,F} \propto C_q^{\frac{1}{q-3}} \mathcal{E}^{-\frac{2}{q-3}} (1 - 3\cos^2\theta)^{-\frac{1}{q-3}}$. If the force changes sign at this zero, its energy barrier height relative to the zero-energy reference state at infinity is defined as $\Delta E_\infty \equiv \mathcal{F}_{tot}(r_F) - \mathcal{F}_{tot}(\infty) = \mathcal{F}_{tot}(r_F) > 0$ whose positivity is ensured by polarization-virial theorem, and it scales as $\Delta E_\infty \propto C_q^{-\frac{3}{q-3}} \mathcal{E}^{\frac{2q}{q-3}} (1 - 3\cos^2\theta)^{\frac{q}{q-3}}$. Thus, for the nonretarded London or classical-thermal regime ($q = 6$), $\mathcal{E}_{c,F} \propto r^{-3/2}$, $r_{c,F} \propto \mathcal{E}^{-2/3}$, $\Delta E_\infty \propto \mathcal{E}^4$. For the quantum-retarded Casimir-Polder regime ($q = 7$), $\mathcal{E}_{c,F} \propto r^{-2}$, $r_{c,F} \propto \mathcal{E}^{-1/2}$, $\Delta E_\infty \propto \mathcal{E}^{7/2}$. For an angular scan at fixed $(r, T, \mathcal{E})$, the critical force-reversal angle scales as $\frac{1}{3} - \cos^2\theta_F \propto \mathcal{E}^{-2}$.

*Ordered cancellation at macroscopic scale.*— The preceding theorems apply most directly to localized dipolar response centers. At a more macroscopic scale, the interacting objects possess spatially-distributed polarization and higher multipoles interactions. To test the robustness of ordered cancellation for extended dielectric objects, we then consider two identical homogeneous dielectric spheres of radius $a$, center separation $R$, and strict charge neutrality in a uniform external electrostatic field. Solving the electrostatic boundary-value problem by spherical-harmonic expansion gives the interaction energy after subtraction of the isolated-sphere contribution; the radial force is obtained by differentiation with respect to $R$. Figure 3(a) shows that finite size shifts both zero angles away from their point-dipole values, with the force-zero angle receiving the larger correction. At $R/a = 3$, the finite-sphere cancellation angles (marked by hollow symbols in Fig. 3(a)) are separated by about 1.92 degree. Figure 3(b) shows that the finite-size corrections decrease systematically with increasing $R/a$ and approach the point-dipole limit in the far zone, while satisfying $\theta_E^{\mathrm{FS}} < \theta_F^{\mathrm{FS}}$. The enhanced near-contact correction to the force zero reflects the stronger weighting of short-ranged multipoles under radial differentiation. Thus, the finite-sphere calculation provides a macroscopic benchmark for the robustness of the ordered-cancellation picture. Finite size shifts the locations of the energetic and mechanical zeros, but does not reverse their order. Quantitative experimental tests would require control of residual charge, electrohydrodynamic interactions, dielectric loss, and thermal fluctuations. The radial force could be measured, for example, by optical trapping[22–25], while the corresponding energy zero could be reconstructed by integrating the force.

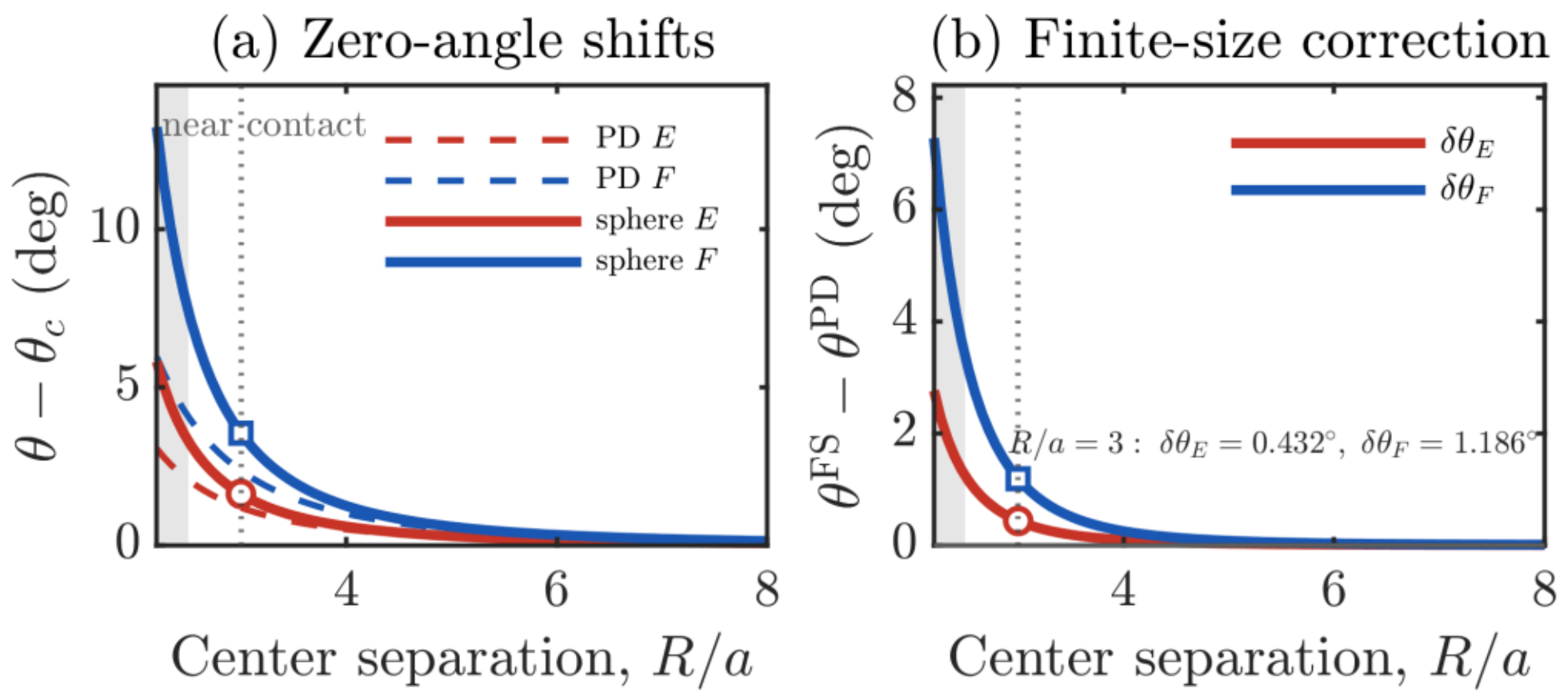

FIG. 3. Finite-size robustness of ordered cancellation, demonstrated by electrostatic multipole calculation for two strictly neutral, identical homogeneous dielectric spheres. The vertical axis is measured relative to the fixed-dipole magic angle $\theta_c = arccos(1/\sqrt{3})$. (a) Energy- and force-zero angle shifts, $\theta_E - \theta_c$ and $\theta_F - \theta_c$, as functions of the center separation $R/a$. Dashed curves show the self-consistent point-dipole (PD) results, while solid curves show the finite-sphere multipole (sphere) results. (b) Finite-size corrections $\delta\theta_E = \theta_E^{\mathrm{FS}} - \theta_E^{\mathrm{PD}}$ and $\delta\theta_F = \theta_F^{\mathrm{FS}} - \theta_F^{\mathrm{PD}}$. The shaded region indicates the near-contact regime $2.2 \leq R/a \leq 2.5$. Open symbols mark the representative point $R/a = 3$, where $\delta\theta_E = 0.432°$ and $\delta\theta_F = 1.186°$. Although finite size and higher multipoles shift both cancellation angles, the ordering $\theta_E^{\mathrm{FS}} < \theta_F^{\mathrm{FS}}$ is preserved.

*Unifying literature insights and discussions.* — The energetic and virial ingredients of induced-dipole interactions are already present in literature[1,3]. Our theorems unite them in a single hierarchy: fixed-source energies and forces, a positive polarization-feedback remainder, a physical virial inequality, monotonicity of the cubically weighted interaction, and ordered cancellation. The results from Anderson and Martin [2] provide an independent check of the static mean-sector theorem. In their notation, let $m_0 = \alpha_{AM} E_0$ denote the isolated induced dipole and $m$ the self-consistent dipole of the longitudinal pair. Using their longitudinal energy-balance expressions, one obtains $V_{AM} - \frac{rF_{r,AM}}{3} = \frac{(m-m_0)^2}{\alpha_{AM}} \geq 0$, which is precisely the longitudinal scalar specialization of our general theorems.

The Phys. Rev. Lett. exchange and subsequent analysis[14–19] focused primarily on the origin of the leading field-dependent term and on limiting field orientations. The present work instead determines the mechanical constraints on the complete sum of the self-consistent mean and equilibrium-fluctuation sectors. We obtain explicit angle-resolved interaction energies and radial forces by combining the closed-form self-consistent mean response with the retarded fluctuation contribution. The leading sign-changing force is the electrostatic force between field-induced mean dipoles, whereas the equilibrium fluctuation contribution remains attractive in the present linear Gaussian setting. Mutual polarization produces an additional attractive correction and shifts the cancellation condition away from the fixed-dipole magic angle.

In the general quantum harmonic oscillator realization, the familiar magic angle is therefore only the weak-coupling limit of two distinct observables: the complete energy-zero angle and the complete force-zero angle. Our polarization-virial theorem grounds this separation as an explicit consequence of positive interaction free energy retained at the force cancellation point. Accordingly, three notions must be distinguished: leading-term cancellation, complete energy cancellation, and complete force cancellation. Although the leading $r^{-3}$ interaction cancels at the conventional magic angle, the mutual-polarization residual begins at order $r^{-6}$. The conventional magic angle must therefore not be identified with either the complete energy-zero angle or the complete force-zero angle.

*Conclusion.—* Closely related to the recent intriguing phenomenon of force-reversal, we reveal two polarization-virial theorems, unified under an entropy-like geometric structure of cubically weighted interaction energy, for describing general field-induced polarization interactions. These theorems cover arbitrary stable linear electric or magnetic dipole assemblies, or reciprocal passive dipole pairs under the stated conditions. They show that energetic and mechanical cancellations cannot coincide: force-reversal point retains positive interaction free energy, leading to ordered cancellation. The general angular structure of force-reversal is revealed through an exact realization of the polarization-virial theorems in a quantum harmonic oscillator pair. Analytical complete arbitrary-angle interaction energy and radial force, which recover quantum-perturbative results, show that the conventional dipolar magic angle splits into strictly ordered energy- and force-zero angles; the latter uniquely bounds the radially repulsive sector and carries a positive energy barrier. Our results delineate the previously unresolved geometry of field-induced force reversal in interatomic long-range forces, with the associate scaling laws of critical phenomena providing accessible experimental consequences that could survive across microscopic and macroscopic scales.

*Appendix*: Notes on angle-resolved quantum perturbation calculation

The perturbative origin of these terms is established by an explicit perturbation calculation, which constructs the dipole coupling relative to the isolated field-dressed oscillators as $V_{dd} = V_{00} + V_{10} + V_{01} + V_{11}$. $V_{00}$ contains the two isolated field-induced dipoles, $V_{10}$ and $V_{01}$ contain one fluctuation operator, and $V_{11}$ contains one fluctuation operator on each oscillator. The first-order contribution from $V_{00}$ is the electrostatic energy $\Delta E_{el}^{(1)} = \alpha\mathcal{E}^2 x(1 - 3\cos^2\theta)$. The second-order contributions from $V_{10}$ and $V_{01}$ describe mutual polarization. Summing over the virtual one-quantum states gives $\Delta E_{pol}^{(2)} = -\alpha\mathcal{E}^2 x^2(1 + 3\cos^2\theta)$, where the explicit factors of $\hbar$ cancel between the transition matrix elements and excitation-energy denominators. In contrast, the $V_{11}$ term couples the ground state to states with one quantum on each oscillator and gives the source-independent London contribution $\Delta E_L^{(2)} = -\frac{3}{4}\hbar\omega_0 x^2$. Consequently, the weak-coupling expansion of the exact interaction is $\Delta E_{nr} = \alpha\mathcal{E}^2[x(1 - 3\cos^2\theta) - x^2(1 + 3\cos^2\theta) + O(x^3)] - \frac{3}{4}\hbar\omega_0 x^2 + O(\hbar\omega_0 x^4)$. Thus the exact Gaussian solution reproduces, term by term, the electrostatic, mutual-polarization, and London sectors obtained from angle-resolved quantum perturbation theory.